\documentclass[prl,twocolumn,amsmath,amssymb,floatfix,letter]{revtex4}
\usepackage[dvips]{graphicx}
\usepackage{color,array,dcolumn}
\usepackage{amsmath}
\usepackage{amssymb}

\begin{document}

\title{Quantum-mechanically enhanced water flow in sub-nanometer carbon nanotubes}

\author{Alberto Ambrosetti$^{1}$\footnote[1]{Corresponding author. Email:alberto.ambrosetti@unipd.it}, Giorgio Palermo$^{1,}$\footnote[2]{Current address: Chaire de Simulation \'{a} l’Echelle Atomique (CSEA), Ecole Polytechnique F\`{e}d\`{e}rale de Lausanne (EPFL), CH-1015 Lausanne, Switzerland} and Pier Luigi Silvestrelli$^1$}

\affiliation{$^1$Dipartimento di Fisica e Astronomia, Universit\`{a} degli Studi di Padova, via Marzolo 8, \textsl{35131}, Padova, Italy}

\begin{abstract}
Water-flow in carbon nanotubes (CNT's) starkly contradicts classical fluid mechanics, with
permeabilities that can exceed no-slip Haagen-Poiseuille predictions
by two to five orders of magnitude.
Semi-classical molecular dynamics accounts for enhanced flow-rates, that are attributed to curvature-dependent
lattice mismatch.
However, the steeper permeability-enhancement observed experimentally at $\sim$nm-size radii remains poorly
understood, and suggests  emergence of puzzling non-classical mechanisms.
Here we address water-CNT friction from a quantum-mechanical perspective, in terms of 
water-energy loss upon phonon excitation. We find that combined weak water-phonon coupling and selection rules hinder water-CNT scattering, 
providing effective protection to water superflow, whereas comparison with a semiclassical theory evidences a 
friction increase that can exceed the quantum-mechanical prediction by more than two orders of magnitude.
Quasi-frictionless flow up to sub-nm  CNT's opens new pathways towards  
minimally-invasive trans-membrane cellular injections, single-water fluidics  and efficient water filtration.

\end{abstract}

%\keywords{carbon nanotubes | water | permeability | superflow }

\maketitle

%%%%%%%%%%%%%%%%%%%%%%%%%%%%%%%%%%%%%%%%%%%%%%%%%%%%%%%%%%%%%%%%%%%%%%%%%%%%%%%%%%%%%%%%%%%%%%%%%%%%%%%%%%%%%%%%%%%

\section{Introduction}
Classical macroscale fluidics~\cite{classic} unambiguously indicates increase of interface flow resistance in  
pipes with smaller radii, owing to larger surface-to-volume ratio.  However, when water is confined to the nm scale, 
ordered low-dimensional phases~\cite{2dice,waterslide,falk} can emerge, and
experimental evidence often contrasts with macroscopic laws.  Exceptionally-high water-flow rates
were observed~\cite{majumder,holt,whitby} in carbon nanotubes (CNT's), with permeabilities
that steeply increase~\cite{secchi} at shorter radii,
exceeding no-slip Haagen-Poiseuille~\cite{secchi,mattia,kannam} predictions by about two orders
of magnitude already at the $\sim$10 nm scale. 
CNT technology promises transformative impact in selective nanoscale fluid-transport,  
surpassing the extraordinary flow-efficiency of cellular membranes, and enabling~\cite{kalra} 
fast, energy-efficient water filtration. Water purification/desalinization devices 
were also proposed~\cite{michaelides} in order to contrast increasing severe shortages of clean water supplies.

Atomistic simulations based on semi-classical potentials could 
predict~\cite{hummer,falk,kannam,mattia,aluru} enhanced water flow rates in CNT's, in qualitative agreement with experiments.
The reduced water-CNT friction was attributed to emergent quasi one-dimensional (1D) water ice~\cite{hummer} 
phases, which imply curvature-induced incommensurability~\cite{falk} with the CNT lattice. Low energetic barriers were independently found by
ab-initio theories~\cite{waterslide,tocci} upon lateral displacement (sliding) of water on 2D graphene~\cite{graphene}.
However, the steep permeability enhancement observed experimentally
at small CNT radii could not be adequately reproduced~\cite{secchi,kannam} so far by semi-classical
atomistic simulations. Experimentally-determined slip lengths~\cite{secchi} far exceed semiclassical~\cite{kannam} predictions, being
already $\sim$10 times larger at $\sim$10 nm. 
%Quantum effects are accordingly expected~\cite{kannam} to acquire increasing
%relevance when CNT confinement approaches the $\sim1$ nm scale. 
Quantum effects may thus acquire increasing relevance when the CNT confinement approaches the $\sim1$ nm scale, but 
their actual role and basic mechanisms remain presently unexplored.
The  electronic structure is also expected to be pivotal in water flow, considering that the
permeability of crystallographically analogous BN nanotubes~\cite{secchi} is much smaller than that of CNT's.

Here we go beyond semi-classical models by introducing a fully quantum-mechanical description of water flow through sub-nanometer-scale CNT's.
Our approach begins with a first-principle assessment of structures and energy landscapes, and then exploits
Fermi's golden rule to account for the energy-loss rate that water undergoes  by scattering with the CNT  phonons.
Water-CNT interactions exhibit non-trivial variability and are heavily suppressed in the presence of CNT metallicity. 
%Moreover, friction is regulated by discrete quantum-mechanical energy transfer between water and CNT  phonons. 
Quasi-vanishing scattering rates, dictated by extremely weak water-phonon coupling and emergent selection rules 
ultimately determine effective quantum mechanical protection of the superflow. 
A "classicization"of our quantum model, relying on a coupled Newtonian dynamics of water and CNT
vibrational modes can enhance friction forces by more than two orders of magnitude. The comparison between quantum and classical models 
accordingly confirms the inherent quantum mechanical nature of  water superflow.

\section{Methods}
We will base our theory
on a first-principle description of the relevant structures and effective water-CNT potentials.
The relevant parameters underlying our quantum mechanical theory are directly computed by density functional theory (DFT). DFT calculations   
are carried out exploiting the Quantum Espresso~\cite{QE} simulation package, based on the semi-local Perdew Burke 
Ernzerhof~\cite{PBE} (PBE) exchange-correlation functional, augmented by D2~\cite{d2} dispersion corrections.
%However, since the main physical effects presented 
%hereafter are attributed to electrostatics and Pauli-repulsion, limited dispersion
%effects on water flow are expected.
Exploiting the introduction of ultrasoft pseudopotentials, electronic wavefunctions are expanded on plane waves with an energy cutoff of 30 Ry. 
The convergence of the  potential energy surface experienced by a water molecule within the (5,5) CNT with respect to the adopted cutoff was explicitly 
checked introducing a 50 Ry cutoff for comparison. The deviation between the
energetic barriers for water flow found comparing 30 Ry and 50 Ry cutoffs amounts to only $\pm\sim$0.04 meV. No qualitative discrepancies are thus  expected
with respect to the present picture. 
We also note that the D2 pairwise approach neglects many-body dispersion terms~\cite{proof,rsscs,science}, which could be of relevance~\cite{carbon,jpcl2} in 
low-dimensional materials. For this reason, in the following we  will further provide an explicit comparison with the many-body dispersion (MBD) 
approach~\cite{rsscs}, which includes both orbital hybridization effects and many-body contributions up to infinite order, 
at an  effective~\cite{proof} random phase approximation (RPA) level. 

All systems are modeled by periodically-repeated simulation cells, whose dimensions are specified in  Tab.~\ref{tab1},
while 20\AA\, vacuum space is enforced orthogonally to the CNT's to minimize spurious periodicity effects.
\begin{table}
    \centering
    \begin{ruledtabular}
    \begin{tabular}{cccc}
           & radius (\AA) & replicas (length) & k-points \\ \hline
           &  &  &\\
(5,5) CNT  & 3.41         & 7 (17.24\AA)      & $4\times 1\times 1$ \\
(7,7) CNT  & 4.76         & 6 (14.77\AA)      & $4\times 1\times 1$ \\
(8,0) CNT  & 3.15         & 4 (17.08\AA)      & $4\times 1\times 1$ \\
(9,0) CNT  & 3.54         & 4 (17.04\AA)      & $4\times 4\times 1$ \\
Graphene   & $\infty$           & $6\times6$ (14.74\AA) & $4\times 4\times 1$ \\
    \end{tabular}
    \end{ruledtabular}
    \caption{Features and settings for the computation of equilibrium energetics and sliding potentials.
Different replicas were adopted, depending on the size of the unit cell. For CNT's, the unit cell coresponds to the minimum
CNT segment which can be periodically repeated along the CNT axis, surrounded by vacuum along orthogonal directions, as defined
in the text.}
    \label{tab1}
\end{table}

Integration over the Brillouin zone is performed by a regular mesh of k-points, and the convention $\hbar=1$ is adopted.

\section{Preliminary structural analysis}
In order to enforce flow of individual water molecules through narrow CNT's, we primarily investigate a (5,5)  
CNT, having a radius of 3.41\AA. This radius is compatible with the optimal water-graphene adsorption distance (amounting to 3.10\AA, measured from the -closest- O atom),
thereby implying weak Pauli repulsion between water and CNT walls.  
The selected CNT has metallic character (namely, zero electronic band-gap at the Fermi level). 
Comparison with (8,0) and (9,0) CNT's having similar radii (3.15\AA, and 3.54\AA, respectively), but 
finite electronic band-gap is also provided in order to specifically assess the role of 
delocalized electrons.

\subsection{Water monomer}
As a preliminary step we analyze the energetics of water confined within the (5,5) CNT. Binding energies are defined as $E_{\rm bind}=E_{\rm tot}-E_{\rm A}-E_{\rm B}$, where $E_{\rm tot}$ is the energy of the total interacting systems, while $E_{\rm A,B}$ are the energies of 
the individual fragments (A,B).
The binding energy of a single water molecule inside the CNT supercell amounts to -396 meV 
(7 replicas of the CNT unit cell, with a total length of 17.24\AA\, are introduced
to minimize fictitious water-water electrostatics). 
The presence of a water monomer within the CNT is thus energetically favorable.
We underline, however, that bulk water may not spontaneously tend to penetrate in the CNT, due to the loss of stabilizing water-water interactions.

In its optimal conformation the water molecule is in the center of the tube, with its dipole moment aligned 
with the CNT longitudinal axis, and a similar configuration is found in the finite-gap (8,0) CNT. 
This contrasts with 2D free-standing graphene, where the water dipole orients 
orthogonally~\cite{jpcc-1} to the surface.

\begin{figure}
%%{\vskip 1.3cm}
\includegraphics[width=8.6cm]{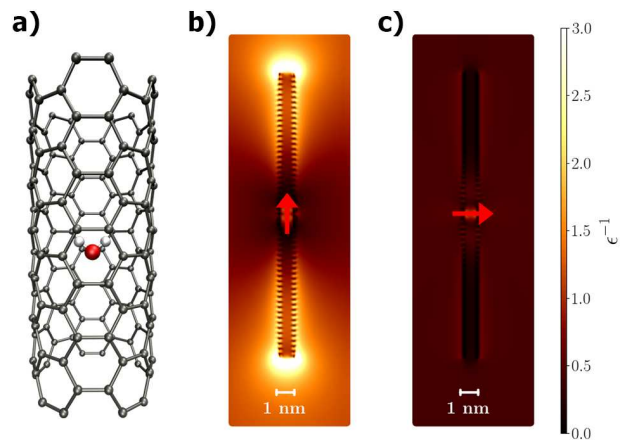}
\caption{(5,5) CNT -- {\bf a)} The water dipole is aligned with the CNT axis in its optimal configuration. The ratio $\epsilon^{-1}$ between the total and bare (i.e. in the absence of CNT) electric-field moduli induced by a unitary dipole in the center of the CNT, is plotted for both longitudinal ({\bf b)}) and normal ({\bf c)}) dipole orientation. The CNT amplifies (suppresses) longitudinal (orthogonal) dipole fields. }
\label{fig1}
\end{figure}

\subsection{Electrostatic analysis}
To further understand the physical mechanism underlying the CNT/dipole alignment we
perform a polarization analysis based on the self-consistent screening~\cite{rsscs}
(SCS) method. Within SCS one maps atomic polarizabilities onto a set of Drude 
oscillators coupled at the dipole level, finally obtaining the screened dipole response of the CNT within the random 
phase approximation~\cite{rsscs}. This allows to compute the total electric field
induced by an internal dipole. Since the SCS implementation is non-periodic, we adopt here an extended 
yet finite (5,5) CNT (similar results are found on other CNT's), having a total length of 134.2\AA.
SCS predicts slight electric field amplification (see Fig.~\ref{fig1}) when the dipole is parallel
to the CNT. The electric field is instead suppressed when the dipole is oriented orthogonally. We recall that electrostatic induction 
grows with the magnitude of the electric field, and can effectively lower the energy of the system. 
Since induction is naturally captured by DFT, the geometrical relaxation of water is thus governed by sterical effects 
(minimization of Pauli repulsion), and electrostatic interactions.

\subsection{Water clusters}
\begin{figure}
%%{\vskip 1.3cm}
\includegraphics[width=8.6cm]{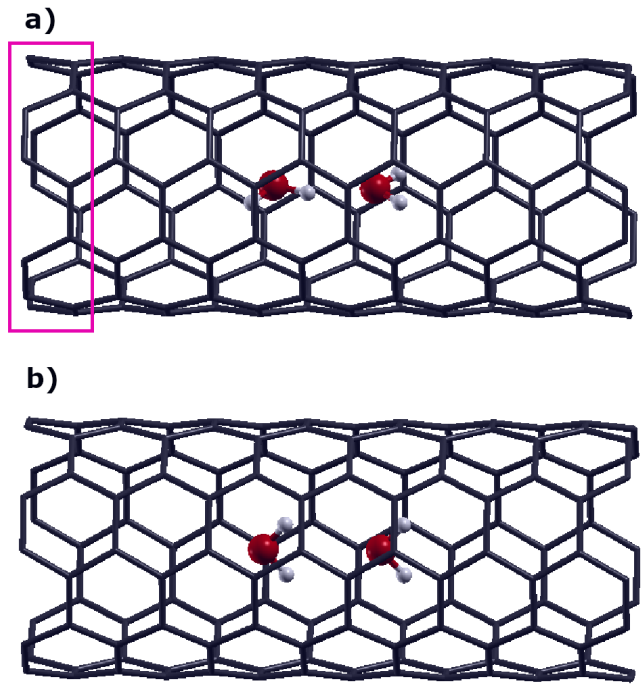}
\caption{Water dimer in the (5,5) CNT -- {\bf a)} H-bonded conformation - the CNT unit cell is indicated by the magenta box. {\bf b)} Conformation with parallel dipoles.}
\label{fig1-2}
\end{figure}

While our focus here is on the flow of individual water molecules, it is also relevant to 
investigate the interaction between multiple molecules confined within the CNT. In the absence of confinement, the water dimer is known 
to form a H-bond (whose energetics amounts to -0.23 eV), and complex ice structures 
can arise~\cite{waterslide,2dice} at interfaces. However, the small (5,5) CNT radius sterically limits
configurational freedom. By structural relaxation we reveal that in the CNT water
can still form H-bonds (see Fig.~\ref{fig1-2}), compatibly with Ref.~\cite{falk}. Upon confinement, the water-water
binding energy becomes -0.27 eV, so that the CNT produces effective H-bond stabilization. 
We note, however, that an alternative stable 
configuration exists, where the two waters are $\sim$3.1\AA\, apart from each other, and are characterized by 
approximately-parallel dipoles, pointing towards the CNT longitudinal axis. This configuration is unfavored with
respect to H-bonding by only 0.14 eV. In the presence of suitable (longitudinally-oriented) 
electric fields that could be introduced to enhance the flow, this
configuration is thus expected to undergo stabilization.
We also observe that while the H-bonded dimer is stabilized by the CNT,  H-bonded trimers can also 
be formed; but under quasi-1D confinement the energetic gain of the second H-bond is only -0.21 eV. 
This should be compared to -0.32 eV, found for the isolated trimer with two H-bonds only. 
In practice, confinement prevents formation of the most stable triple-H-bonded trimer configuration, 
and the bond enhancement encountered in the dimer does not apply to longer water chains.
We also note that no periodic  purely H-bonded 1D structure was detected in the CNT.
Considering that H-bonding implies limited energy improvement, and that highly
selective configurational arrangement is required for the formation of H-bonds, one expects that at low
density or in the presence of suitable electric fields,  {\it longitudinal-dipole} configurations
should largely contribute to the water flow. In fact, alignment of water dipoles (Fig.~\ref{fig1-2}-b) remains viable
for arbitrarily large water clusters.

In passing, we remark the role of exact exchange for the correct description of hydrogen bonding~\cite{hbond1,hbond2,hbond3}.
In fact, it is known that dispersion-corrected PBE tends to overestimate the H-bond stability~\cite{hbond1}. This implies that
multiple hydrogen-bond formation should be further hindered, favoring alignment of (unbound) single water dipoles.
On the other hand, we recall that the electrostatic interactions relevant to our system are  reliably described by (semi-)local DFT
due to the inbuilt self-consistent treatment of electron density and total electrostatic potential.

\section{Energy barriers for water flow}
We address at this point the energetic barriers that water at low density needs to overcome when flowing 
through the CNT, due to the presence of the surrounding C atoms. A water monomer in its optimal configuration is rigidly translated along the longitudinal axis, in order to 
sample the corrugations of the potential energy $V_{\rm w}$ computed by DFT.
We note that $V_{\rm w}$ ultimately depends on the position of water ($\mathbf{r}_{\rm W}$) and on the Cartesian coordinates (${\mathbf R}$) of the C atoms in the CNT.
For the moment we will consider the CNT atoms as frozen in their equilibrium position ($\bar{{\mathbf R}}$), and will explicit the dependence of $V_{\rm w}$ on the longitudinal coordinate
of water ($x_{\rm W}$).

Within the metallic (5,5) CNT, the energy corrugations $\Delta V_{\rm w}$ with respect to $x_{\rm W}$ are quasi negligible, 
amounting to $\sim\pm$0.35 meV, with sinusoidal oscillations (see Fig.~\ref{fig2}). 
Computation of the same barrier at the PBE+MBD level with 50 Ry cutoff yields an even smaller value ($\sim \pm 0.27$ meV).
The adopted computational setup can thus be regarded as prudential.
Conversely, in the gapped (8,0) CNT corrugations become $\sim 10^2$ times larger (i.e. $\Delta V_{\rm w}\sim\pm$30 meV). 
We note that the radius of the (8,0) CNT is $\sim$0.2\AA\, smaller than the (5,5). 
However, in the small-gap (9,0) CNT having larger radius %, whose radius is 3.54 \AA\, 
$\Delta V_{\rm w}$  still amounts to $\sim\pm$8 meV. Given that (5,5) and (8,0)-(9,0) CNT's have comparable radii, metallic electrons 
are proven  pivotal  in determining the relevant energetic landscape. As a confirmation, $\Delta V_{\rm w}$ in the larger, 
metallic (7,7) CNT still amounts to $\sim\pm$0.45 meV: delocalized $\pi_z$ electrons  are known to  constitute a thin layer of 
charge~\cite{dobson-es} which effectively screens the details of the underlying atomic lattice.
We observe in this regard that both theoretical simulations~\cite{waterslide} and 
experiments~\cite{graf-lub} evidence high graphite lubricity in the presence
of humidity, providing independent confirmation of low water-graphene friction.

For comparison, we additionally computed the energetic corrugation by sliding a single water molecule 
in the optimal conformation (with dipole moment orthogonal to the surface) on 2D free-standing 
graphene. The corrugation in this 
case amounts to  $\sim\pm$6 meV. However, $\Delta V_{\rm w}\sim\pm$3 meV
 is found when the water dipole is oriented parallel to the surface, while 
keeping the water-graphene distance unchanged. 
Hence, both curvature and water orientation~\cite{aluru} contribute to the energy landscape, in addition to band structure.  
Notably, only marginally larger corrugation ($\sim\pm$0.8 meV) is found by sliding a H-bonded water 
dimer through the same CNT. % in spite of the more complex structure.

\section{Quantum theory of water flow}
The extremely-low sliding barriers found in the (5,5) CNT already suggest quasi-frictionless water
flow from a semi-classical perspective. 
However, in order to address the problem from a truly quantum mechanical point of view, we should 
consider that the energy transfer between water and CNT implies the excitation of phonons.

We model here water as a single quantum particle (neglecting internal degrees of freedom), subject to the weakly corrugated potential $V_{\rm w}$ described above.  The Hamiltonian associated to a water molecule in the CNT environment is 
\begin{equation}
H_{\rm w}=T_{\rm w}+V_{\rm w} \,,
\end{equation}
where $T_{\rm w}$ is the kinetic energy of water, and $V_{\rm w}$ is the water-CNT potential deescribed above.
We expect that the quantum effects~\cite{prot-trans,prot-trans2} conventionally addressed in water (proton transfer) should not be crucial to the present analysis: key quantum mechanisms regard here the energy transfer between water and CNT.

Ideally, when the CNT atomic coordinates are frozen, water can indefinitely occupy any eigenstate $|\psi_{\rm w}^i\rangle$ (such that $H_{\rm w}|\psi_{\rm w}^i\rangle = E_{\rm w}^i |\psi_{\rm w}^i\rangle$). 
However, quantum mechanical CNT vibrations (phonons) must be taken into account.
These are effectively described by the quadratic Hamiltonian 
\begin{equation}
H_{\rm ph}=\sum_{i} \frac{P^2_{R_i}}{2 m_{\rm C}} + \sum_{i, j} \left( \frac{\partial_{R_i}\partial_{R_j} E_{\rm CNT}}{2} \right)_{R=\overline{R}}\, \delta {R_i} \delta {R_j} \,.
\end{equation}
Here $i,j$ run over both atomic indices and Cartesian components,  $P_{R_i}=-i \partial_{R_i}$, % $(x,y,z)$, 
and $\delta R_i $ represents the displacement from the equilibrium position $\bar{R}_i$.
$E_{\rm CNT}$ is the energy of the CNT, obtained by DFT for a given conformation $\mathbf{R}$, and $m_{\rm C}$ is the 
mass of a C atom.  Diagonalization of $H_{\rm ph}$
yields the normal vibrational modes of the systems, whose coordinates and frequencies will be 
indicated as $\tilde{R}_i=S_{ij} R_j$, and $\omega_i$, respectively. Repeated indices are contracted, and $S$ is the matrix of transformation from cartesian atomic coordinates to collective phonon coordinates.

Atomic vibrations imply variability of the coordinates $R_i$. When these displacements are small, $V_{\rm w}$ varies as:
\begin{equation}
\label{vwater}
V_{\rm w} (\mathbf{r}_{\rm W},{\mathbf{R}})=V_{\rm w} (\mathbf{r}_{\rm W},{\mathbf{\bar{R}}}) + \sum_{i} \partial_{R_i} V_{\rm w}(\mathbf{r}_{\rm W},{\mathbf{R}}) \vert_{\mathbf{R}=\mathbf{\overline{R}}} \, \delta {R_i}+O(\delta\mathbf{R}^2 )\,.
\end{equation}
The second term to the right introduces an effective water-phonon coupling: in fact, the operators $\delta R_i$ can either excite or annihilate single CNT phonons. 
Confined water can thus scatter against the CNT, transferring part of its energy to the CNT lattice upon phonon excitation.
The corresponding energy loss in flowing water can ultimately be regarded as a quantum mechanical source of friction. 

To physically address the water flow process, we exploit Fermi's golden rule, which expresses the transition probability
per unit time ($\Gamma^i$) relative to water in the initial state $\psi_{\rm w}^i$ as: 
\begin{equation}
\label{fermi}
\Gamma^i = 2 \pi \sum_{
\substack{
j\neq i \\
m
}
} 
\vert \langle \psi_{\rm w}^i \vert \partial_{R_l} V_{\rm w} \vert \psi_{\rm w}^j \rangle  \langle 0_m  \vert \delta R_l  \vert 1_m \rangle|^2 \delta(E_{\rm w}^i - E_{\rm w}^j - \omega_m) \,.
\end{equation}
Here the repeated index $l$ is contracted, $| 1_m \rangle$ indicates first excitation of the $m$-th phonon, and phononic ground state
is assumed in the initial setup.
The last factor to the right ensures conservation of the total energy within the scattering process: the energy lost by water must be converted here into phonon excitation.

Water eigenstates could be obtained by numerical diagonalization of $H_{\rm w}$, assuming
that the motion is one-dimensional (1D), and that $V_{\rm w}$ is sinusoidal (as empirically confirmed by DFT calculations). 
However, owing to the extremely small corrugation 
$\Delta V_{\rm w}$, both $\psi_{\rm w}^i$ and $E_{\rm w}^i$ are well approximated  by 
free-particle
eigenstates: the energetic scale of phonon excitations is between $\sim50$ and $\sim680$ times larger than $\Delta V_{\rm w}$. 
The $i$-th eigenstate is thus associated to a plane wave with 
1D wavevector $k_i$, and energy $E_{\rm w}^i=k_i^2/(2 m_{\rm w})$ ($m_{\rm w}$ being the water mass).
Exploiting the algebra of quantum harmonic oscillators we note that $\delta \tilde{R}_l $ can only introduce a single excitation on the ground state, and 
$\langle 0_m  | \delta \tilde{R}_l  | 1_m \rangle = \delta_{lm}/\sqrt{2 m_{\rm C} \omega_m}$. 
Moreover, by DFT calculations one consistently observes that $\partial_{R_l} V_{\rm w}$ can be recast in the form
$C_l+A_l\,\cos(k_1 x_{\rm W} +\phi_l)$, being $C_l$, $A_l$, $\phi_l$ mode-dependent constants (see Appendix),
and $k_1=2\pi/L$, where $L$ is the unit cell length. Hence, considering that water should lose energy during the
scattering process, the term $\langle \psi_{\rm w}^i \vert \partial_{R_l} V_{\rm w} \vert \psi_{\rm w}^j \rangle$
implies the following selection rule on momenta: $k_j=k_i-k_1$ (as straightforwardly obtained by Fourier analysis).

In order to keep the problem tractable from the computational point of view, we restricted here to
phonon modes within the CNT unit cell. However, the approach can be extended to arbitrary supercells,
(or $k$-points) where analogous results are expected, although $\partial_{R_l} V_{\rm w}$ could be associated in this case to different wavevectors.
The six lowest-frequency modes are also removed in order to account for the lack of translational and rotational freedom in fixed CNT's.
At sufficiently low temperature one expects that the phononic ground state approximation
is justified (although  presence of excitations does not qualitatively alter the present conclusions), 
and single phonon excitations  provide the leading contribution to Eq.~\eqref{fermi}. 

Simultaneous fulfillment of energy conservation and selection of momenta imply that scattering and excitation of the $i$-th phonon 
can take place only if  $k_i=m_{\rm w} \omega_i /k_1 + k_1/2$.
This relation indicates that for each phonon mode there exists only a single initial state 
$\psi_i$ (with wavevector $k_i$) compatible with the transition. This constraint might 
be {\it softened} by lattice distortions and many-body effects, but it is expected to hold in general at least in 
approximate form, since it directly follows from symmetry and energy-conservation criteria.
%We further remark that phonon modes beyond the $\Gamma$-point in the CNT unit cell
%could be characterized by small $k_1$ (slow oscillation of the $V_{\rm w}$ derivative). 
%At low $k_1$ phonon frequencies~\cite{phon} can decrease, but water-phonon couplings are also expected to be weakened 
%due to the smooth geometry distortions.
To provide a more exhaustive picture, CNT phonons\cite{phon} in the relevant limit of vanishing wavevectors will be qualitatively addressed in Appendix C.
Even considering phonon dispersion, we expect that our $\Gamma$-point description of the scattering
process should already provide a correct qualitative description of the phenomenon.

Considering the above analysis, after straightforward algebra Eq.~\eqref{fermi} can finally be recast as
\begin{equation}
\label{scatt-prob}
\Gamma^i = \frac{2 \pi }{8 m_{\rm C} \omega_i} \vert \sum_n A_n S_{ni}  \vert^2 \,,
\end{equation}
where, again, $k_i$ is univocally associated to $\omega_i$, as seen above.

\section{Results}
\begin{figure}
%%{\vskip 1.3cm}
\includegraphics[width=8.5cm]{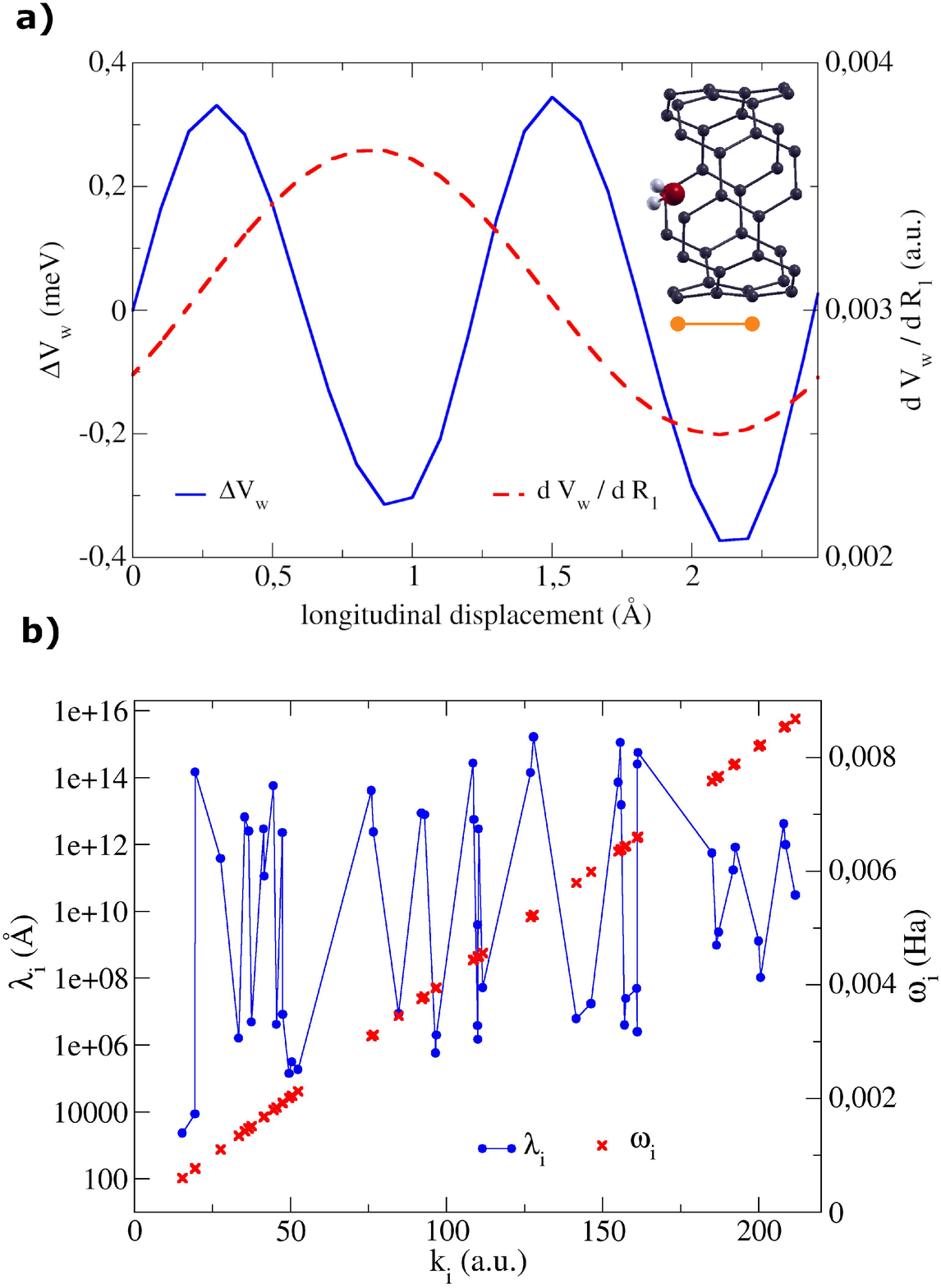}
\caption{(5,5) CNT: %\bf{a)} the water dipole is aligned with the CNT axis in its optimal configuration.
 {\bf a)} Potential corrugation $\Delta V_{\rm w}$ and potential derivative with respect to a selected C atom, plotted along the CNT unit cell (longitudinal coordinate). The orange segment in inset depicts a unit-cell displacement. Both curves can be approximated as sinusoidal, although having different periods. Analogous conclusions are drawn for all derivatives. {\bf b)} Estimated mean free path $\lambda_i$ for each wave vector $k_i$ compatible with the scattering. The phonon frequencies $\omega_i$ corresponding to each $k_i$ are given for reference.}
\label{fig2}
\end{figure}

Starting from Eq.~\eqref{scatt-prob} we will now estimate the mean free path that water can travel before scattering. 
We note that $(\Gamma^i)^{-1}$ can be interpreted as the time interval $\Delta t$ intercurring between
subsequent scattering processes. Considering a wavepacket centered around a given wavevector $k$, this can be associated to the group velocity $v_k=k/m_{\rm w}$. 
We can thus interpret $\lambda_i=(\Gamma^i)^{-1} k_i/m_{\rm w}$ as a measure of the effective mean free path. 
From Fig.~\ref{fig2} we observe that $\lambda_i$ in (5,5) CNT's roughly varies between $\sim10^5$ and 
$\sim10^{15}$ \AA, with the exception of two low-frequency outliers (corresponding to low wavevectors),
where $\lambda_i\sim 10^3-10^4$ \AA.

Considering that a traveling wavepacket is a superposition of many wavevectors,
the effective mean free path $\lambda$ should be in between the aforementioned ranges.
This means that water could travel for many micrometers before scattering. %(i.e. well beyond the $\mu$m scale),
Hence, tubes with nanoscale length are expected to support water superflow, encountering almost no resistance.
On the other hand, we observe that  water speeds reported in the  water flow experiments of Ref.~\cite{secchi}
were in the 1-10$\mu$m/s range. This is much smaller than the minimum group velocity compatible with scattering,
considered here. According to our model, essentially no scattering with unit-cell phonons is permitted at the corresponding (small)
wavevector $k_i$: in practice, a very slow water molecule does not carry sufficient kinetic energy to excite the phonons under consideration.  However,
scattering with other phonons associated to low wavectors (or large supercells) cannot be excluded.
In Appendix C we will see that constraints on permitted scattering transitions should also be expected for the phonons with vanishing frequency~\cite{phon},
that can be found in CNT's in the limit of vanishing wavevector. Hence, the present conclusions are expected to be general.

Within our model we  can also  qualitatively estimate permeability enhancement factors, i.e.
ratios between quantum mechanical and no-slip Haagen-Poiseuille permeabilities, which can be compared to
available experimental values.
We first note that the average quantum mechanical friction force $F_{\rm i,QM}$ acting on water can be
estimated as the energy loss per unit length  $F_{\rm i,QM}\sim \Delta E_{\rm i} / \Delta L_{\rm i} $. Since single phonon excitation
with energy transfer $\Delta E_{\rm i}\sim \omega_m$ occurs on average after $\Delta L_{\rm i}\sim \lambda_i$ ,
one finally has
\begin{equation}
\label{f-av}
F_{\rm i,QM}\sim \frac{\omega_i}{\lambda_i} %= \frac{m_{\rm w} \omega_m \Gamma^i}{ k_i}= \frac{2 \pi m_{\rm w} }{8 m_{\rm C} k_i } \vert \sum_n A_n S_{nm}  \vert^2\,.
\end{equation}
%Hence, the somewhat larger mean free paths obtained at low $\omega_m$ will still be associated to a low
%friction force, although lower friction should be expected at higher water velocity if the couplings remain unvaried.
If $S$ is the transversal  section (or area) of the CNT, the pressure demanded for water flow is
\begin{equation}
\label{dp}
\Delta P_i=\frac{F_{\rm i, QM}}{S}=\frac{\omega_i}{\lambda_i S} \,.
\end{equation}
The permeability $\kappa_p^i$ is  defined in terms of the volumetric water flux $Q^i$, by the relation  $\kappa_p^i=Q^i L/(\Delta P^i S)$, where
$L$ is the CNT length.
By exploiting Eq.~\eqref{dp} we thus find $\kappa_p^i=Q^i L \lambda_i/\omega_i$.
Semi-classically, the water flux can be written as $Q^i= S v^i$, hence
\begin{equation}
\kappa_p^i=\frac{ k_i S L \lambda_i }{m_{\rm W}\omega_i}\,.
\end{equation}
If we consider now the standard no-slip Haagen-Poiseuille~\cite{classic} permeability $\kappa_{\rm HP}=R^2/(8\mu)$ ($\mu$ being the viscosity
of water), we find the enhancement factor
\begin{equation}
\label{enhanc}
\frac{\kappa_p^i}{\kappa_{\rm HP}}=\frac{k_i S L \lambda_i }{m_{\rm W}\omega_m} \frac{8\mu}{R^2}\,.
\end{equation}
We now take a model (5,5) CNT with L=5 nm.
Exploiting the dynamical viscosity of water at 300K
one obtains that $\kappa_p^i/\kappa_{\rm HP}$ can very between $\sim 10^6$ and $\sim 10^{17}$, depending
on the specific $\lambda_i$ (and associated $k_i$).
%(considering $10^5$\AA\, $< \lambda_i < 10^{14}$\AA), 
The range of these enhancement factors can explain the large experimental estimates~\cite{holt,qin} (i.e. $\sim10^3-10^4$) found in CNT's having $\sim$1 nm 
radius.  In fact, exceedingly-high enhancement factors will be effectively hidden in 
experiments by wavepacket convolution and by energy losses at CNT edges: the passage of water from the CNT to outer space and viceversa 
was found to produce relevant dissipation~\cite{ebrahimi}  (neglected here), in analogy with macrofluidics.
We thus expect that higher enhancement factors could be experimentally measured in longer CNT's, where edge effects become less relevant.
We also note that, given the small water-CNT coupling and the quasi-negligible energy transfer,
the harmonic approximation to lattice vibrations (phonons) appears well justified.
Anharmonic effects are expected to slightly renormalize phonon energies without qualitatively altering the present picture. It is however
expedient to recall that higher-level (computationally demanding) approaches, such as as self-consistent phonons~\cite{scphon} or ring polymer molecular
dynamics~\cite{ring} could describe vibrational modes beyond the harmonic assumption.

While our theory was developed for very narrow CNTs, and does not provide a priori a general law for the scaling of $\lambda^i$, we expect that
$\lambda^i$ should decrease at larger $R$. The steep  $\kappa_p^i/\kappa_{\rm HP}$ enhancement
observed~\cite{secchi} in experiments at short $R$ can thus be connected to a fast decay of
$\lambda^i$, in qualitative agreement with the present analysis. We also recall that additional "classical" dissipation
mechanisms will likely arise beyond $R\sim 1$ nm, as predicted~\cite{hummer,falk,kannam,mattia,aluru} by molecular dynamics.

Concerning finite-gap (8,0) CNT's,  somewhat shorter $\lambda_i$ are expected, given the larger energy barriers $\Delta V_{\rm w}$. We note, 
however, that while larger $\Delta V_{\rm w}$ can imply more significant deviation
from the free-particle description of water, the parameters $A_n$ have larger impact on
 scattering effects, as they contribute quadratically to $\Gamma^i$.
Numerical tests reveal that while $\Delta V_{\rm w}$ in the (9,0) CNT is more than an order of magnitude 
larger than in (5,5), the oscillations of the potential derivatives $\partial_{R_i} V_{\rm w}$ 
have comparable magnitude (see Appendix B). 
Nevertheless, owing to the different structure, more complex oscillatory patterns are identified in 
(9,0) CNT, implying the coexistence of higher Fourier components, which can actively contribute to 
the scattering process (see Eq.~\eqref{fermi}).
%We also note that $\Gamma^i$ is inversely proportional to $\omega_i$, so that low phonon frequencies will be associated to smaller mean free paths. 

While the present theory is meant to address sub-nanometer CNT's, it still supports 
qualitative interpretation of the lower permeability observed in larger CNT's.
Semiclassical molecular dynamics  suggests~\cite{aluru,falk,hummer} that water can organize into
ordered structures when confined in narrow CNT's, so that the flow is effectively optimized.
This picture is compatible with our theory, which reveals that narrower CNT's exhibit a
preference for longitudinal dipole orientation, or can induce quasi-1D ordered phases. 
At larger radii, where semiclassical theories become increasingly accurate,  
disordered liquid structures can emerge, and water-CNT's interactions slowly approach the 2D graphene limit. 
Water molecules have much larger configurational freedom in that case, and can redistribute energy between
CNT and other water molecules, limiting the effectiveness of the above selection rules.
Also, by analogy with
continuum elastic theory, the walls of larger nanotubes are expected to be easily deformed:
identical forces applied to larger elastic structures -or springs- imply larger deformations.
Hence, the phonon spectrum will be enriched by additional degrees of freedom,
with~\cite{phon}  low-frequency~\cite{phon} modes. All these mechanisms should effectively enhance 
$\Gamma^i$.
Finally, the significant variability of experimental permeability
estimates at comparable CNT radii~\cite{kannam} can be understood in terms of electronic structure and water orientation.

\section{Semiclassical approximation}
So far, we demonstrated that surprisingly large mean free paths and permeability enhancement factors are predicted by our quantum mechanical model,
compatibly with the experimental evidence. Now, we will further demonstrate how quantum mechanics truly influences friction forces by a 
direct comparison between a classical and quantum mechanical analysis of the same physical model. 

We introduce a "classicization" of the quantum model exploited so far,
taking inspiration from the Frenkel-Korontova~\cite{frenkel} approach. 
Water  is modeled here as a classical particle moving longitudinally through the CNT, subject to the 
same effective potential $V_{\rm W}$ (Fig.~\ref{fig2}) exploited so far.
As in the quantum mechanical approach, $V_{\rm w}$ depends on the ionic coordinates $\mathbf{R}$,
and we account for this dependence up to first-order in the displacements from the equilibrium positions $\bar{\mathbf{R}}$,
according to Eq.~\eqref{vwater}. CNT lattice vibrations are now treated as classical harmonic oscillators (with the same collective coordinates $\tilde{R}_i$ as before), and
the following set of equations eventually regulates the motion of the whole system:
\begin{eqnarray}
\label{dyn}
m_{\rm w}\partial_t^2\mathbf{r}_{\rm w} = -\nabla_{r_{\rm w}}V_{\rm w}({\mathbf{R}},\mathbf{r}_{\rm w})|_{\mathbf{R}=\bar{\mathbf{R}}} 
-\nabla_{r_{\rm w}}\sum_i \partial_{R_i} V_{\rm w}({\mathbf{R}},\mathbf{r}_{\rm w})|_{\mathbf{R}=\bar{\mathbf{R}}} \delta R_i\,, \nonumber \\
m_{\rm ph}\partial_t^2 \tilde{R}_i = -m_{\rm ph}\omega_i^2 \delta \tilde{R}_i  -\partial_{\tilde{R}_i} V_{\rm w}({\mathbf{R}},\mathbf{r}_{\rm w})|_{\mathbf{R}=\bar{\mathbf{R}}}\,,
\end{eqnarray}
where $m_{\rm ph}$ is the phonon mass.
The  equations of motion (restricted to 1D) are integrated exploiting Verlet's algorithm, imposing that ions are initially ($t=0$)
at rest in their equilibrium configuration, and that the initial speed of water $v_0=k/m_{\rm w}$ is parallel to the CNT.
The time interval for integration is fixed by adopting $10^6$ steps within the simulation cell (of 2.46\AA) This is
sufficient to ensure convergence for any initial speed $v_0$ considered here.
%The energy transfer between water and ions is estimated as the sum of the kinetic and harmonic potential energies acquired by
%the ions at a given time step. %Constant terms in $\partial_{R_i} V_{\rm w}$ are neglected, since they do not couple water and ionic motion, and can be attributed to a tiny residual stress which has little relevance in this context.
Following the trajectory of water up to three cell lengths, one empirically observes quasi-linear growth of the energy transfer between water and CNT with respect to time,  
(perturbed by small periodic modulations) for all adopted velocities. It is thus appropriate to estimate the average energy transfer per
CNT unit length  $F_{\rm k,C}\sim \Delta E_{\rm C} / \Delta L_{\rm C} $  in analogy to Eq.~\eqref{f-av}, where
$L_{\rm C}$ and $E_{\rm C}$ are the  classical path length and energy loss of the traveling water molecule.

\begin{figure}
%%{\vskip 1.3cm}
\includegraphics[width=8.5cm]{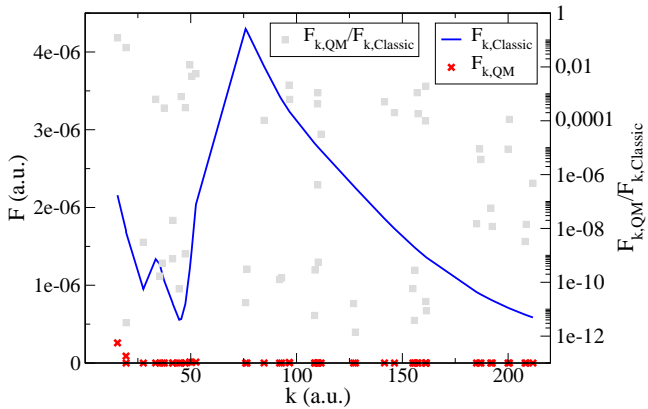}
\caption{(5,5) CNT: comparison between effective friction forces predicted by the quantum
mechanical and classical approaches. The ratio between quantum and classical forces 
 is also reported for direct comparison. 
Within the (semi)classical model $k$ determines the initial water speed $v_0$ according to the relation $v_0=k/m_{\rm w}$.}
\label{fig3}
\end{figure}

Since all degrees of freedom are now 
classical, no selection rules emerge. Hence, given any initial speed of the water molecule, all vibrational
modes can be simultaneously excited, with continuous energy transfer. No dissipative term is introduced for CNT vibrational modes to avoid arbitrary
parametrization. In Fig.~\ref{fig3} we observe that the quantum mechanical forces 
in the (5,5) CNT are much smaller than semiclassical predictions, by a factor
$\sim10^{-2}-10^{-12}$, depending on $k$ (with the only exception of the first two outliers, where  the ratio amounts to $\sim10^{-1}$). This directly demonstrates that quantum effects heavily suppress friction forces. 
We also note that force ratios in Fig.~\ref{fig3} are roughly distributed in two separate groups, in analogy to $\lambda_i$. In fact, there exists some phonon modes that are
essentially uncoupled to water (lower points), where force ratios are extremely small (and mean free paths are extremely large). 
The geometries of these phonons are consistently found to be antisymmetric.
In the other group (upper points, with finite coupling) we observe instead that the ratio between classical and quantum
forces tends to decrease with respect to $k$. This suggests that the quantum mechanical suppression should become more effective at larger speed.
To rationalize the large difference between quantum and classical results we recall that, quantum-mechanically:
{\it i)} only a single phonon can be activated  at the corresponding $k_i$, {\it ii)} energy can only be transferred by discrete amounts,
and {\it iii)} selection rules due to symmetry and energy conservation must be respected. 

We finally remark that (semi-)classical approaches will be intrinsically unable to capture the major quantum-mechanical suppression of friction forces
reported in this work, regardless of their accuracy. On the other hand, quantum mechanical friction is so small, that a slight variation of the adopted model, or a 
marginal improvement of the underpinning first-principle method is not expected to significantly alter the overall qualitative picture.

\section{Conclusions}
In conclusion, a quantum mechanical description is demanded to qualitatively rationalize the giant 
permeability  observed experimentally in sub-nm wide CNT's.  Fermi's golden rule is exploited here, 
avoiding computationally demanding path-integral molecular dynamics simulations, and possible biases~\cite{kannam} related to thermostats.
%, without the need to introduce possible biases~\cite{kannam} related to thermostats, as conventonally done in molecular dynamics.
The interaction induced by the CNT on confined water can vary depending on water orientation, electronic structure 
and curvature, and exhibits quasi-vanishing, sub-meV sliding barriers in narrow metallic CNT's. 
Quantum mechanically, interface friction is regulated by discrete water-CNT energy transfer due to phonon excitation.
A combination of  weak water-phonon coupling, discrete energy transfer
and selection rules is found to hinder scattering phenomena, providing quantum mechanical protection to the superflow of water, whereas (semi-)classical approaches can severely 
increase water-CNT friction forces.

We anticipate that application of external electric fields may be exploited in order to fine-tune 
the superflow, due to the interplay with longitudinally aligned water dipoles.
Water superflow could be exploited for optimization of nanofluidic phenomena, including controlled
and minimally-invasive injection of few water molecules through cellular membranes, water transport in artificial photosynthetic 
systems~\cite{photo}, or energy-efficient water filtration~\cite{michaelides,kalra} devices that could effectively contrast
increasingly severe shortages of clean-water supplies. We also suggest that similar quantum mechanical mechanisms
may also interest graphene/graphite surfaces, where strong lubricity was consistently observed~\cite{graf-lub} in the presence of water.

%\begin{suppinfo}
%Details about first-principle calculations, the SCS model, and the semi-classical approach are provided as supplementary material.
%\end{suppinfo}

%%%%%%%%%%%%%%%%%%%%%%%%%%%%%%%%%%%%%%%%%%%%%%%%%%%%%%%%%%%%%%%%%%%%%%%%%%%%%%%%%%%%%%%%%%

\section{Acknowledgments}
A.A. and P.L.S. acknowledge funding from Cassa di Risparmio di Padova e Rovigo (CARIPARO) - grant EngvdW. A.A. acknowledges funding from Cassa di Risparmio di Padova e Rovigo (CARIPARO) - grant Synergy.
%\end{acknowledgements}
%%%%%%%%%%%%%%%%%%%%%%%%%%%%%%%%%%%%%%%%%%%%%%%%%%%%%%%%%%%%%%%%%%%%%%%%%%%%%%%%%%%%%%%%%%
%\bibliography{literature.bib}
%\bibliographystyle{apsrevtex4}

\section{Appendix A: Self-consistent screening approach}
The idea behind the self-consistent screening (SCS) approach
is to map the dipole response of $N$ atoms into a collection of QHOs, obtaining a discretized
expression for the electrodynamic screening equation.
Given $N$ atoms, we thus define $A^{\rm SCS}$ and $A^0$ as the interacting and bare ($3N\times3N$) polarizability tensors, respectively, where $A^0_{lm}=\delta_{lm} \alpha_l$ (bare atomic polarizabilities are assumed to be local). Hence, one obtains:
\begin{equation}
A^{\mathrm{SCS}}_{lm}(i \omega)=A^{\mathrm{0}}_{lm}(i \omega)-A^{\mathrm{0}}_{lp}(i \omega) T_{p q} A^{\mathrm{SCS}}_{qm}(i \omega)\,.
\label{eq:scs_atoms}
\end{equation}
Here indices run over both atomic and Cartesian coordinates, and are contracted when repeated.  $T_{p q}$ is the dipole-dipole interaction tensor, defined by double derivatives of the Coulomb interaction $T_{pq}=\partial_p \partial_q v_{Coul}$ (with respect to coordinates $p$ and $q$).
The Coulomb potential is damped at short range according to the prescriptions of Ref.~\cite{rsscs},  to account for charge overlap between atoms. Free atom polarizabilities are obtained within the Tkatchenko-Scheffler~\cite{ts} approach, and rescaled proportionally to the Hirshfeld volume ratio between atoms in the CNT and free atoms, which is fixed here to 0.87, according to previous~\cite{science} calculations.
Solution of Eq.~\eqref{eq:scs_atoms} can be obtained by matrix diagonalization, so that the dipole tensor $A^{\mathrm{SCS}}$  can be
computed exactly.
The electric field at position $\mathbf{r}'$ (with Cartesian index $k$), due to a permanent dipole
$\mathbf{d}$ placed at $\mathbf{r}''$ (with Cartesian index $l$) is then given by the sum of the bare field and that induced by the polarization of all $N$ atoms, according to the SCS dipolar response.
This is expressed as:
\begin{equation}
E_{k} = -T^{r',r''}_{kl} d_{l} + T^{r',r_m}_{km} A^{\rm SCS}_{mj} T^{r_j, r''}_{jl} d_{l}\,,
\end{equation}
where superscripts indicate the coordinates connected by dipole tensors.

\section{Appendix B: Potential derivatives}
In order to compute the derivatives of the $V_{\rm w} $ potential for the (5,5) CNT with respect to the positions of the C atoms
within unit cell, the simulation was reduced to 2 supercells. The contributions relative to periodically
repeated atoms were summed up, so that the correct periodic functional forms were eventually obtained. In the (9,0) CNT, instead, the unit
cell was sufficiently extended to accommodate a water molecule.

\begin{figure}
{\vskip 1.3cm}
\includegraphics[width=8.2cm]{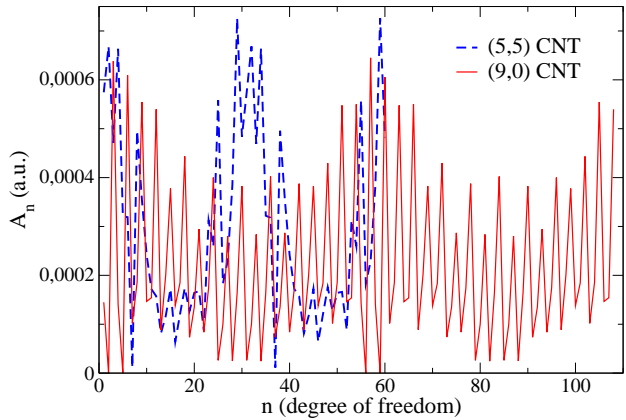}
\caption{ $A_n$ parameters --defined as half the difference between maximum and minimum values of
$\partial_{R_n} V_{\rm w}$ within the cell-- for all degrees of freedom $n$, computed in both (5,5) and (9,0) CNT's. }
\label{fig-potder}
\end{figure}

The parameters $A_n$, describing the amplitude of the potential derivatives for (5,5) and (9,0) CNT's are reported in Fig.~\ref{fig-potder} for comparison. The six lowest-frequency phonons were discarded, given
their relation to macroscopic translations and rotations, which are absent in CNT's with fixed position.
We note that the periodicity of the potential derivatives may slightly vary in different supercells, including for instance finite Fourier components whose wavevector is shifted by integer
multiples of $k_1$, compatibly with umklapp processes. 

\section{Appendix C: Phonons in the limit of vanishing wavevector}
We recall that our ($\Gamma$-point) unit cell model predicts that scattering can only occur if the constraint $k_i=m_{\rm w} \omega_i /k_1 + k_1/2$ is satisfied.
Here $k_1$ is the wavevector associated to the derivatives of the potential  $V_{\rm w}$ with respect to the ionic coordinates, $k_i$ is the wavevector which describes water in the initial state,
and $\omega_i $ is a phonon frequency at the $\Gamma$ point.
However, when phonons up to infinite cells are considered, their frequencies $\omega'_i$ can substantially vary, and the derivatives of $V_{\rm w}$ may also
include components with different wavevectors $k'_1$ (we recall that $k_1$ is relative to the unit cell, and it characterizes the CNT lattice).

First of all, we note that $V_{\rm W}$ depends on the difference between $\mathbf{r}_{\rm W}$ and each of the ionic coordinates $\mathbf{R}$. Hence, by  symmetry, 
a rigid translation of all coordinates $\mathbf{R}$ along the CNT axis (which actually corresponds to a  longitudinal phonon with zero wavevector) is equivalent to an opposite translation of
the water molecule. Hence, for this translational mode one obtains a potential derivative which can be computed as $-\partial_{x_{\rm W}} V_{\rm w}$. 
Considering the sinusoidal form of $V_{\rm w}$, as from Fig.\ref{fig2}, the derivative can be computed analytically, 
yielding $-2A_0 k_1 \sin(2k_1 x_{\rm W})$, where $A_0\sim$0.35 meV. According to this formula, this mode implies very small water-phonon coupling (of the order of $10^{-5}$ a.u.).
In addition, the potential derivative is still associated to a finite wavevector (i.e. $2k_1$). One may thus speculate that even in the case of slowly-varying phonons, a finite wavevector
will effectively control the potential derivatives.

In this case, even if the phonon frequency is vanishing in the zero wavevector limit, the above constraint should be recast as $k_i=  k_1$.  
This still implies a direct correspondence between $k_i$ and the allowed phonon excitations:  only those 
phonons with $\omega'_i=0$ can be excited by water with initial wavevector $k_i=k_1$, and analogous calculations can be performed for higher-frequency phonons. 

We should now consider that other Fourier components, whose wavevector is differing by integer multiple of $k_1$ can emerge in the potential derivatives (for instance considering larger supercells),
compatibly with umklapp scattering processes.
In case the potential derivatives have a finite Fourier component with vanishing wavevector $k'_1\rightarrow 0$, one faces two possibilities: either the phonon frequency $\omega'_i$ is finite, or 
$\omega'_i \rightarrow 0$. If $\omega'_i$ is finite, then $k_i\sim m_{\rm w} \omega'_i /k'_1 $ tends to diverge when $k'_1\rightarrow 0$. Hence, only very fast water molecules can scatter against these modes. Instead, if 
$\omega'_i \rightarrow 0$
one should consider that the phonon spectra\cite{phon,phon2} for (5,5) CNT exhibits a few quasi-linear bands at vanishing wavevectors, so that $\omega'_i\sim c_i k'_1$, where $c_i$ are
proportionality constants which can be extrapolated from the phonon dispersion. From Refs.~\cite{phon,phon2} we estimate that $\lim_{k'_1\rightarrow 0} m_{\rm w} \omega'_i /k'_1$
is of the order of $10^2$ a.u., which still implies a rather large $k_i$ in order to enable scattering.

Last, we further consider the case of very slow water molecules, where the adopted free-particle spectrum  may provide a poor description of the actual dispersion of water eigenmodes. 
In this case,
water eigenfunctions associated to a small wavevector $k_i $ could possess additional Fourier components with wavevectors  $k_i+n k_1$ (where $n$ is a integer number), 
due to the presence of the periodic potential $V_{\rm W}$ (the effects of $V_{\rm W}$ are instead negligible at high $k_i$). 
The scattering in this case should take into account possible {\it shifts} in momenta by integer multiples of $k_1$. Hence, scattering could be permitted even with low $k_i$ (and low water speed) if the
potential derivatives include a Fourier component with arbitrarily small wavevector $k'_1$. However, only a limited number of {\it shifts} 
will provide a sizable contribution, which limits again the availability of allowed transitions. Basically, the same qualitative conclusions are expected to hold also for umklapp processes,
where, again, phonon momentum is conserved only up to a reciprocal lattice vector.

In conclusion, the limited number of permitted transitions implies quantum mechanical suppression of friction forces in all limiting cases, including both high water velocities, and
small speeds, in qualitative agreement with experiments~\cite{secchi}.  

%in all the limiting cases considered here, water can move through the CNT with low friction even at the low $k_i$ values compatible with experimental velocities~\cite{secchi}. We also expect 
%that even at much higher (yet finite) speeds, quantum mechanics should suppress friction forces with respect to classical mechanics, as seen in the unit cell analysis.

%%%%%%%%%%%%%%%%%%%%%%%%%%%%%%%%%%%%%%%%%%%%%%%%%%%%%%%%%%%%%%%%%%%%%%%%%%%%%%%%%%%%%%%%%%%%%%%%%%%%%
\end{document}